\documentclass[aps,prl,twocolumn,
superscriptaddress]{revtex4}
\usepackage{amssymb}
\usepackage{graphicx}
\usepackage{amsmath}
\usepackage{color}

\begin{document}

\title{Universal Logarithmic Scrambling in Many Body Localization }
\author{Yu Chen}
\email{yuchen.physics@cnu.edu.cn}
\affiliation{Center for Theoretical Physics, Department of Physics, Capital Normal University, Beijing, 100048, China}

\begin{abstract}
Out of time ordered correlator (OTOC) is recently introduced as a powerful diagnose for quantum chaos. To go beyond, here we present an analytical solution of OTOC for a non-chaotic many body localized (MBL) system, showing distinct feature from quantum chaos and Anderson localization (AL). 
The OTOC is found to fall only if the nearest distance between the two operators being shorter than $\xi\ln t$, where $\xi$ is dimensionless localization length. Thereafter, we found an universal power law decay of OTOC as $2^{-\xi\ln t}$, implying an universal logarithmic growth of second R\'{e}nyi entropy, where $\xi$ plays the role of information scrambling rate. 
A relation between butterfly velocity and scrambling rate is found.

\end{abstract}

\maketitle
\emph{Introduction} 
Out-of-time-ordered correlation (OTOC) function is recently introduced as a very powerful diagnose for quantum chaos\cite{OTOC01,OTOC02,OTOC03,OTOC04,OTOC05,OTOC06}. Originally, this is first applying to AdS/CFT systems, where gravitational duality exists. These systems are proved to be extremely chaotic and reached the upper bound of chaos\cite{OTOC06}. An analog of Lyapunov exponent is found just parallel to nonlinear response study of disorder superconductors\cite{LO} with a four point correlator for early time,
\begin{eqnarray}
C(t)=\left\langle [\hat{W}(t),\hat{V}]^\dag  [\hat{W}(t),\hat{V}]\right\rangle_\beta\sim e^{\lambda_L t},\nonumber
\end{eqnarray}
where $\langle \cdot\rangle_\beta$ means ${\rm Tr}(e^{-\beta \hat{H}}\cdot)$ as thermal average, $\beta=1/k_BT$ being inverse temperature. $\hat{W}$ and $\hat{V}$ are two general operators. The OTO correlator is defined as the out-of-time-ordered part of $C(t)$,
\begin{eqnarray}
F(t)=\left\langle \hat{W}^\dag(t)\hat{V}^\dag(0)\hat{W}(t)\hat{V}(0)\right\rangle_\beta,
\end{eqnarray}
hence $C(t)=2(1-{\rm Re}F)$. 
Here $\lambda_L$ is the analog of Lyapunov exponent, which is orginally proposed in classical systems describing how exponential large is the deviation of trajectory could accumulate in time after a small perturbation. Interestingly, a connection between this dynamical instability and entropy is discovered as Kolmogorov-Sinai entropy\cite{KS_entropy}, which interprets Lyapunov exponent as entropy rate, bridging two seemly irrelevant stories. 
Surprisingly, a quantum version analog of ``KS entropy" also exists in OTOC, where $\lambda_L$ is interpreted as information scrambling rate, indicting the formation of mutual information that cannot be extracted from local measurements\cite{Qi}. Therefore the OTOC seems to carry important messages about information spreading, quantum thermalization and so on, which are of tremendous interest today.

Retrospecting the definition of OTOC, we find it quite general. But our knowledge for OTOC is still limited. Most of present studies of OTOC are centered on fast scrambled systems with gravity duality like SYK model\cite{SY, Sachdev, Kitaev} or systems with large number of conserved quantities like rational CFT\cite{Yingfei, Alvaro} . A natural question then is to ask how will it behave in more general respect, what if the system is not so fast scrambled? 
These curiosities lead us to go beyond. Here we explore how OTOC behaves in a MBL system\cite{Anderson58, Altshuler06, Mirlin06}. In a closed system MBL states could not reach their thermal equilibrium after a quantum quench\cite{Huse07, Huse10, Huse15}, which has been proved experimentally in disordered interacting fermions and bosons\cite{Bloch15a, Bloch15b, DeMarco, Bloch16}. A logarithmic growth of entanglement entropy is predicted\cite{Pollman12, Chiara06, Prosen08, Huse14a, Huse14b, Altman13, Altman14, Abanin13a} and time-ordered spectral functions are calculated\cite{Rahul14a,Rahul14b,Rahul15,Rahul16}, MBL state is believed to be slow scrambled and not an efficient heat bath to thermalize itself. 
\begin{figure}[t]
\includegraphics[width=7.5cm]{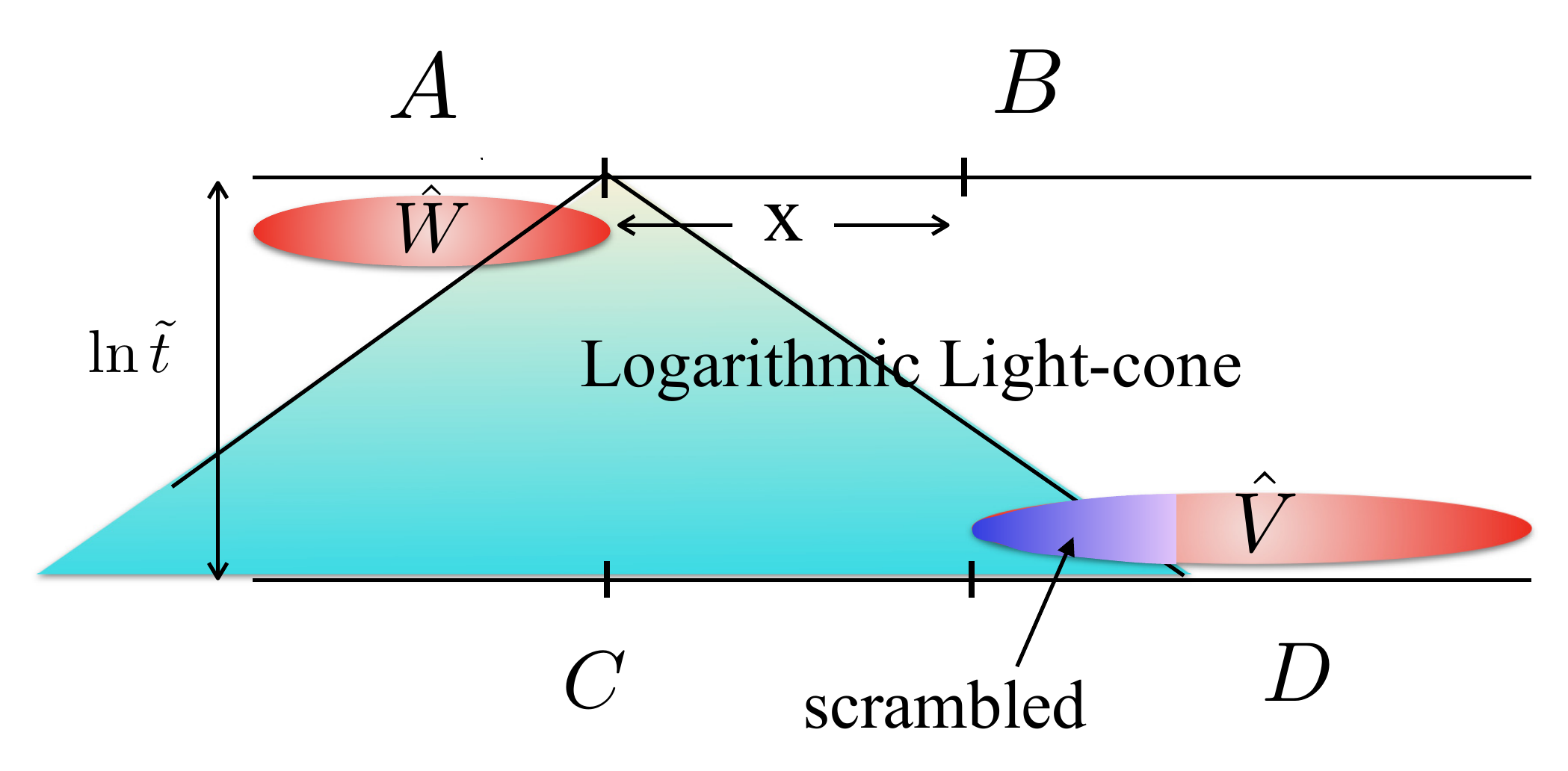}
\caption{Illustration for quantum channel scheme for OTOC. A and D are output and input of the quantum circuit. The evolution time is $\tilde{t}$, shown in logarithmic scale. A logarithmic light-cone is drawn from inner side of A, where D enters the time-like region. Minimal distance between A and D is x.}
\label{LLC}
\end{figure}
So what could OTOC tell us about MBL? Could OTOC give us more information like entanglement spreading? In this Letter, we are going to present an analytical calculation of OTOC in MBL systems with the help of fixed point phenomenological model and try to answer these questions. Our main finding is there exist a typical time $\tau_B$ for OTOC to start falling, defined as $\log \tau_B=x/\xi$, where $x$ is the minimal real space distance between $\hat{W}$ and $\hat{V}$, $\xi$ is the localization length. This is a manifestation of logarithmic version of Lieb-Robinson bound\cite{Lieb-Robinson,LR02,LR03,Burrell07,Stolz12,Abanin14,Brian-LR,LLC02}(Lieb-Robinson bound is a finite velocity bound for information traveling velocity in spin systems). When $\xi\log t>x$, the information begins to be scrambled and the OTOC decays in a power law with an universal exponent $\xi\ln 2$, independent of disorder distribution. Translated into R\'{e}nyi entropy, it implies an universal logarithmic scrambling in MBL. Finally the OTOC reaches its lower bound after $\xi\ln t$ reaches system size.

\emph{Phenomenological Model}
 For simplicity we start from a fixed point effective hamiltonian for MBL state\cite{Abanin13b,Huse15}.
\begin{eqnarray}
\hat{H}=\sum_i h_i \tau_i^z+\sum_{ij} J_{ij}\tau_i^z\tau_j^z+\sum_{ijk}J_{ijk}\tau_i^z\tau_j^z\tau_k^z+\cdots
\end{eqnarray}
where $\tau_i^z$ denote local conserved quantity. Here for simplicity, let us take $\tau_i^z=\sigma_i^z$ where $\sigma_i^z$ is the Pauli matrix in z direction at site i. And we drop all terms with more than two spin interactions, then the hamiltonian is simplified as  $\hat{H}_0=\sum_i h_i \sigma_i^z+\sum_{ij} J_{ij}\sigma_i^z\sigma_j^z$. Here we assume it is in localization region and $J_{ij}$ is a random interaction strength decays exponentially with distance, $J_{ij}=J^0_{ij} e^{-|i-j|/\xi}$, where $\xi$ is dimensionless localization length. $J^0_{ij}$ is random and satisfies $\langle J^0_{ij}\rangle=0$, $\langle (J_{ij}^0)^2\rangle={\cal J}^2$. First we assume the disorder satisfy Poisson distribution $P(J)=e^{-2|J|/{\cal J}}$, later we will compare results between different distributions.

The OTOC we are going to calculate is over two finite size operators with minimal distance $x$, and normalized by $\sqrt{\langle\hat{W}^\dag(t)\hat{V}^\dag(0)\hat{V}(0)\hat{W}(t)\rangle_\beta}$ and $\sqrt{\langle\hat{V}^\dag(0)\hat{W}^\dag(t)\hat{W}(t)\hat{V}(0)\rangle_\beta}$, denoted as $f(t)$. Here we focus only on infinite temperature case $\beta=0$. 
More specifically, we study a one dimensional spin chain and we take one spin chain as input, another for output to generate a quantum circuit as is shown in Fig.~\ref{LLC}, which is proposed in Ref.~\cite{Qi}. Further we divide the input spin chain at time zero as part C and D; the output spin chain at time t as part A and B. Now we require operator $\hat{W}$ is from part A, denoted as $\hat{W}\in\hat{\cal O}_A$ and $\hat{V}$ is from operator restricted in part D, $\hat{V}\in{\cal O}_D$. In a recent pioneer work\cite{Qi}, average over all $\hat{W}$ and $\hat{V}$ for infinite temperature $\beta=0$ gives a simple relation between the OTOC function $f(t)$ and the second Renyi entropy 
\begin{eqnarray}
S_R^{(2)}(t)=-\ln f(t)/\ln2,
\end{eqnarray}
by which we could establish relation between OTOC and information scrambling.

\emph{Analytical Expression for OTOC} First of all, let us assume $\#(A)=M'$ and $\#(D)=M$, both count from the edge, with $\#(A)$ standing for number of sites in part A. Meanwhile the minimal distance between A and D is $x=N+1-M-M'$, where N is total sites number. Let us denote $|n\rangle=|\pm_1\cdots\pm_N\rangle$ as a spin configuration with $\pm_i$ as eigenvalue for $\sigma_i^z$. We consider $\hat{W}=\otimes_{i=N}^{N-(M'-1)}\sigma^{[\mu]}_i$, $\hat{V}=\otimes_{j=1}^{M-1}\sigma^{[\nu]}_j$, where $\sigma^{[\mu]}$,($\mu=0,1,2,3$) are $1$, $\sigma^x$, $\sigma^y$, $\sigma^z$. A short notation is introduced as $\hat{W}=[\mu_N\cdots \mu_{N-M'+1}]$, $\hat{V}=[\nu_1\cdots \nu_{M}]$. 

\begin{figure}[t]
\includegraphics[width=7.8cm]{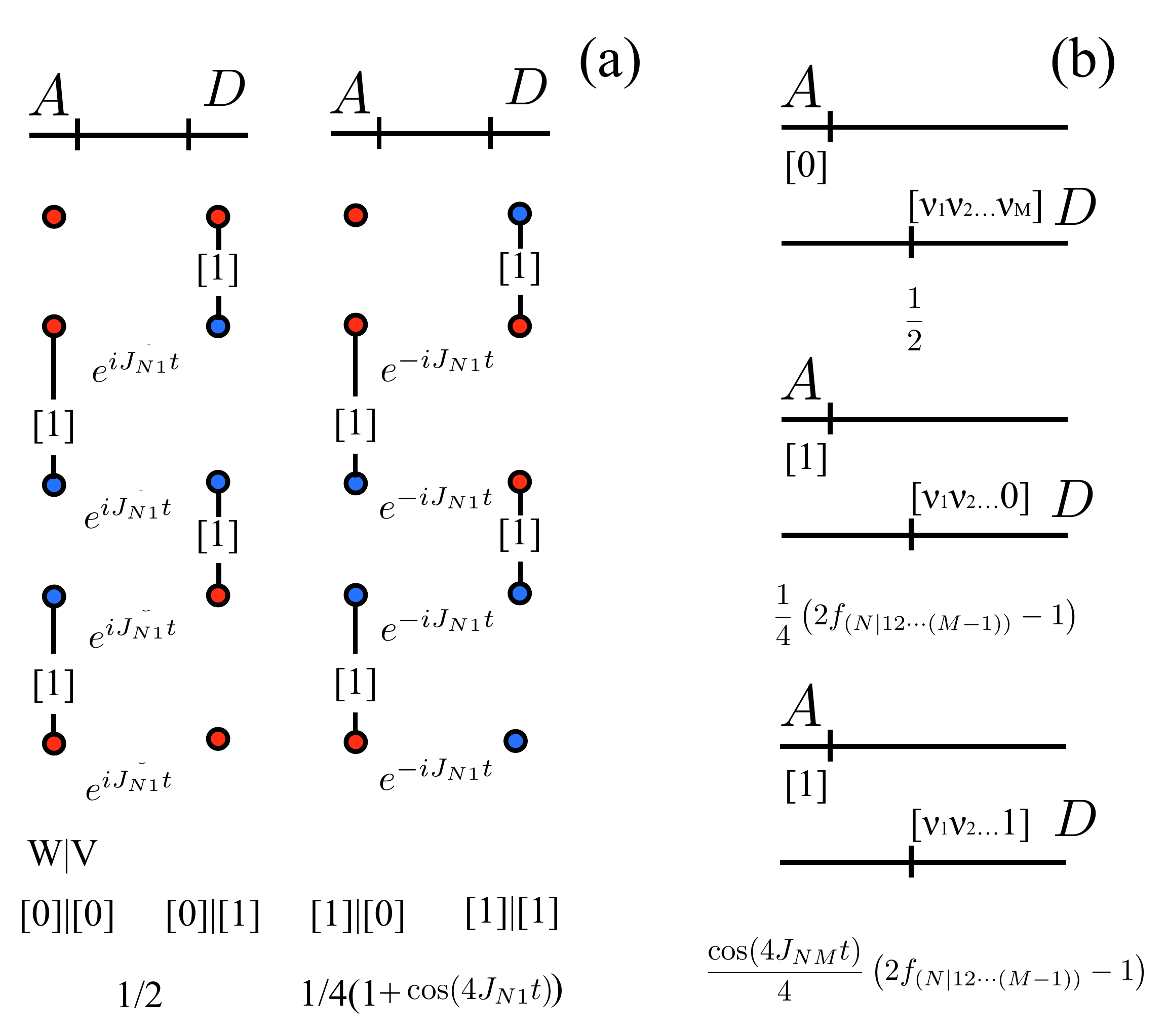}
\caption{(a) one explicit calculation for $\hat{W}=[1]$  and $\hat{V}=[1]$ case.  Red and blue dot are spin up and down. (b) Recursion strategy.}\label{Recursion}
\end{figure}

Now we start from the simplest case $\hat{W}=[\mu_N]$ and $\hat{V}=[\nu_1]$. Instead of calculating the operator evolution, here we calculate the path integral literately by summation over all different paths of spin configurations. Then we find only spin configuration on site $1$ and $N$ are relevant. In Fig.2 (a), an explicit evolution path is shown. and we could find that 
\begin{eqnarray}
f_{(N|1)}(t)=\frac{1}{2}\left(1+\frac{1}{2}(1+\cos(4J_{N1}t))\right),
\end{eqnarray}
where $f_{(A|D)}(t)$ is introduced as notation for OTOC between A and D. 
The only nontrivial channel is when $\hat{W}=[1]$ or $[2]$ while $\hat{V}=[1]$ or $[2]$. For this reason we found that the only difference between these operators are $\sigma^{1,2}$ flip the spin while $\sigma^{0,3}$ does not. Then we can further simplify the notation as $0$ for $\sigma^0$ and $\sigma^3$ and $1$ for $\sigma^1$ and $\sigma^2$. From now on $[0]$ stands for $[0/3]$, $[1]$ stands for $[1/2]$. Now we add site 2 to part D.
Then if $\hat{W}=[0]$, no matter what $\hat{V}$ takes, it is trivial, that make $\frac{1}{2}$. As is shown in Fig.~\ref{Recursion}(b), with $\hat{W}=[1]$, if the new added operator on site 2 is [0], then it is $\frac{1}{2^2}(2f_{(N|1)}-1)$; else if the new added operator in $\hat{V}$ is $[1]$, then the contribution is $\frac{1}{2^2}(2f_{(N|1)}-1)\cos(4J_{N2}t)$, then we can deduct that $2f_{(N|12)}-1=(2f_{(N|1)}-1)(2f_{(N|2)}-1)$. Repeat this recursion, we could get the OTOC for $A=\{N\}$ and $D=\{12\cdots M\}$.
\begin{eqnarray}
f(t)=\frac{1}{2}\left(1+\left(\frac{1}{2}\right)^M\prod_{i=1}^M(1+\cos(4J_{Ni}t))\right)\label{Eq:ND}
 \end{eqnarray}
 
In the same spirit, our recursive step also works for adding sites to A, as long as A and D shares no common site. We have $f_{(N,N-1|D)}(t)=f_{(N|D)}(t)f_{(N-1|D)}(t)$. For $A=\{N(N-1)\cdots (N-M'+1)\}$ and $D=\{12\cdots M\}$, 
\begin{eqnarray}
f(t)=\prod_{j=1}^{M'}\!\left[\frac{1}{2}\!\left(1+\left(\frac{1}{2}\right)^{M}\!\prod_{i=1}^{M}(1+\cos(4J_{(N+1-j)i}t))\right)\!\!\right]\!\!.
\end{eqnarray}
We could also see $\lim_{t\rightarrow\infty}f(t)=2^{-M'}$ for the second case, which satisfy the definition of chaos better. Here we stress that our recursion scheme only works for $M'\leq M$ case. When $M'>M$ we can start from the other half.

\emph{Universal power law decay of OTOC} Now we turn to calculate these expressions more explicitly to extract universal behavior out of it. Let us first calculate $A=\{N\}$ and $D=\{12\cdots M\}$ case. After disorder average over Poission distribution of $J_{ij}^0$ we have $f(t)=\frac{1}{2}(1+2^{-M}\exp(g(t)))$, where
\begin{eqnarray}
g(t)&=&\sum_{i=1}^{M} \ln\left(1+\frac{1}{(8{\cal J} t e^{-(N-i)/\xi})^2+1}\right)\nonumber\\
&\approx&\xi\ln 2\left(\ln\frac{\sin\arctan(8{\cal J}t e^{-(N-M)/\xi})}{\sin\arctan(8{\cal J}t e^{-N/\xi})}\right).
\end{eqnarray}
Here we introduce $\tilde{t}=8{\cal J}t$, and $x=N-M$ being the shortest distance between $\hat{W}$ and $\hat{V}$ in real space. For AL, ${\cal J}=0$, therefore $f(t)=1$. But any finite ${\cal J}$, no matter how small, makes it quite different. For ${\cal J}\neq 0$, immediately we find two time scales, the first one is $\tau_B$ satisfying $\log \tau_B=x/\xi$, while the other one is $\tau_S$ satisfying $\log \tau_S=(x+M)/\xi$. Further we introduce a function $S(x)=\ln\sin\arctan\exp(x)$, then 
\begin{eqnarray}
g(t)=\xi\ln2\left(S(\ln\tilde{t}-\ln\tau_B)-S(\ln\tilde{t}-\ln\tau_S)\right)
\end{eqnarray}
Notice function $S(x)$ satisfy for $x<0$, $S(x)\sim x$; for $x>0$, $S(x)\sim 0$.
Then we could observe that for $\tilde{t}<\tau_B$, $g(t)\sim M\ln 2$, that is $f(t)=1$. While $\tau_B<\tilde{t}<\tau_S$, then $g(t)=\ln 2(M-\xi\ln\tilde{t})$, 
\begin{eqnarray}
f(t)=\frac{1}{2}(1+2^{-\xi\ln \tilde{t}}).
\end{eqnarray}
\begin{figure}[t]
\includegraphics[width=8.6cm]{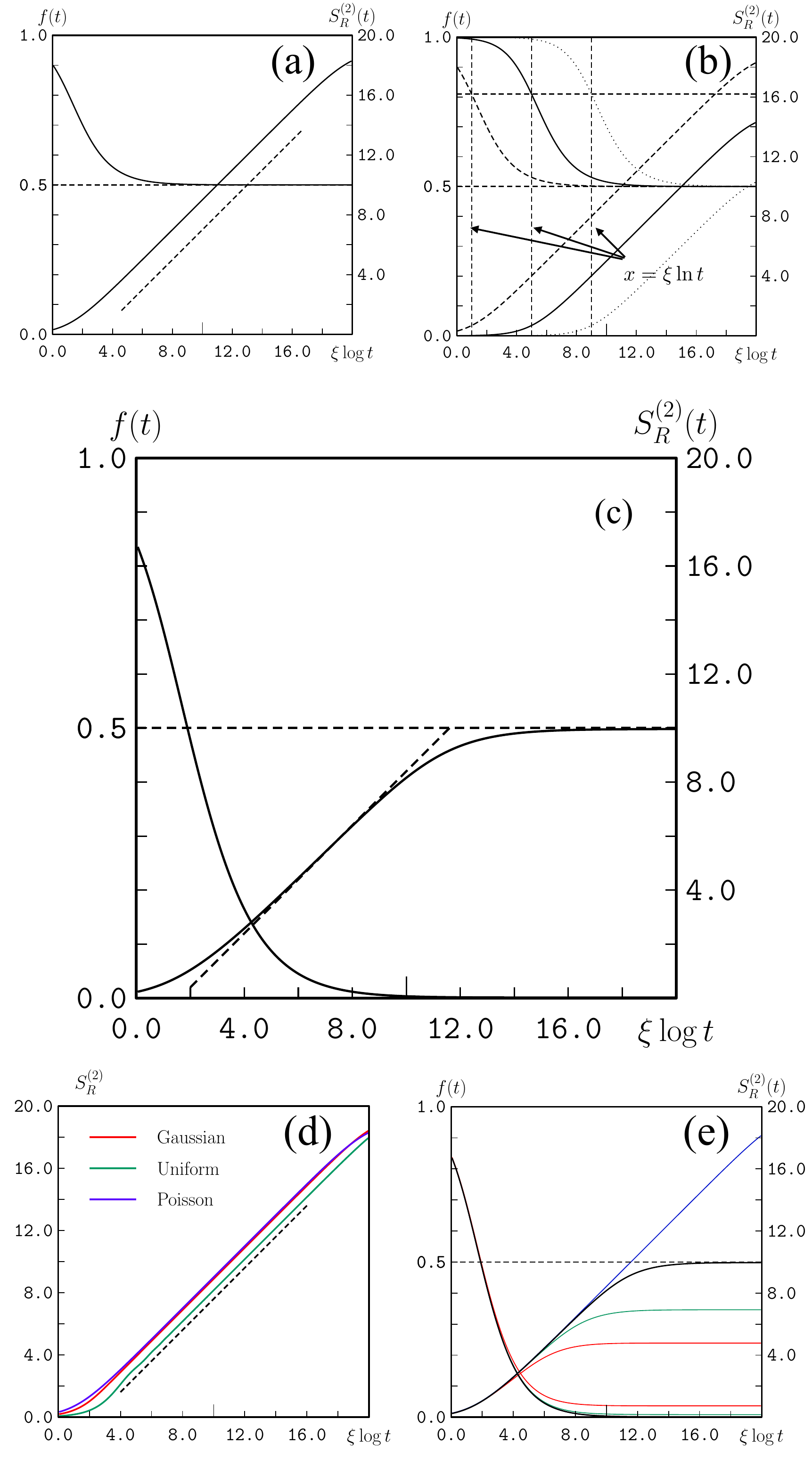}
\caption{In (a) we show OTOC $f(t)$ and second R\'{e}nyi entropy $S_R^{(2)}$ for $(A|D)=(N|12\cdots N-1)$. Size of the system is taken to be $N=20$. The horizontal axis are all taken as $\xi\log t$. (b), we give OTOC for $(A|D)=(N|12\cdots M)$ with M=19,15,11. In (c) we the OTOC $f(t)$ and second R\'{e}nyi entropy $S_R^{(2)}$ for $(A|D)=(N(N-1)\cdots\frac{N}{2}+1|12\cdots \frac{N}{2})$. In (d) we shows how R\'{e}nyi entropy is different with different disorder distribution, the $\log t$ law is robust and the slopes are the same.  (e) we give finite size scaling of the f(t) studied in (c). red for N=10, green for N=14 and blue for N=40. }\label{Fig:OTOC}
\end{figure}
\!\!We also find around $\tau_B$, $f(t)\sim 1-c_0\exp(2(\ln\tilde{t}-x/\xi))$, where $c_0$ is disorder distribution dependent costant.
Compared with chaotic case, $\xi$ is the butterfly velocity\cite{Qi} and this phenomenon meet with the logarithmic Lieb-Robinson bound and logarithmic light-cone\cite{Abanin14,Brian-LR,LLC02} in MBL systems.
Finally when $\tilde{t}>\tau_S$, $g(t)\sim 0$, that is $f(t)=1/2$. Now it becomes clear that the function $S(\ln t-x/\xi)$ has an information interpretation, when $\ln t-x/\xi<0$, that is outside of the logarithmic light-cone, by Lieb-Robinson bound, they are uncorrelated. While $\ln t-x/\xi>0$, the information is scrambled and merge into the whole system, at the same time the R\'{e}nyi entropy begins to grow, in a speed of $\xi$, as shown in Fig.~\ref{Fig:OTOC}(b).  We could see $\xi\ln 2$ is information entropy within localization volume.  Clearly the information been scrambled is proportional to information arrived at the LLC, and this is proportional to $\xi\ln t$. $\xi$ could be understood as the speed of information hit the boundary of this logarithmic light-cone, and everything arrive this boundary will be scrambled. This is how butterfly velocity is connected to scrambling rate.  
For this reason, we will call $\tau_B$ the butterfly arrival time, describing the time for A and D step into each other's LLC. $\tau_S$ is saturate time, describing the time for the whole D passes through the LLC of A. 

One can observe that the the index of power law decay $\xi\ln 2$ has nothing to do with disorder strength ${\cal J}$, and could be interpreted as information scrambling rate proportional to entropy in localization cell, therefore it should be disorder distribution independent. Without a formal proof, we can testify this guess by changing distribution. Here we take the uniform distribution $P(J)=1/(2\sqrt{2}{\cal J})\theta(\sqrt{2}{\cal J}-|J|)$, in the same middle range of time, we have
\begin{eqnarray}
g(t)=\sum_i\ln\left(1+\frac{\sin(\sqrt{2}{\cal J}4t e^{-i/\xi})}{\sqrt{2}{\cal J}4t e^{-i/\xi}}\right)\approx-\xi\ln2\ln\tilde{t}
\end{eqnarray}
with $\tilde{t}=4\sqrt{2}{\cal J}t$. As we expected, only the time unit is changed by disorder distribution, not the power index. More accurate numerical integrations of $g(t)$ for different disorder distribution rescaled to the same time unit has been shown in Fig.~\ref{Fig:OTOC}(d), showing an universal logarithmic growth of second R\'{e}nyi entropy independent of disorder distribution.

Now we turn to a more general case, we take $\#(A)=\#(D)=N/2$ from two side, then 
\begin{eqnarray}
f(t)&=&\prod_{j=1}^{N/2}\left(\frac{1}{2}\left(1+2^{-N/2}\exp(g_j(t))\right)\right)\\
g_j(t)&=&\xi\ln2\left(\ln\frac{\sin\arctan(8{\cal J}t e^{-j/\xi})}{\sin\arctan(8{\cal J}t e^{-(j+N/2)/\xi})}\right)
\end{eqnarray}
$g_j$ describes a process of D entering site $N-j+1$'s LLC. And this process happens for each site in A one by one.  For $\ln\tilde{t}$ being large enough but smaller than $N/2\xi$, we have 
\begin{eqnarray}
f(t)=2^{-1-\frac{1}{2}(\xi\log t)^2}+2^{-\xi\log t}
\end{eqnarray}
While the first term becomes negligible, we could see the universal power $\xi\ln2$. From the connection between R\'{e}nyi entropy and OTOC function, we get
\begin{eqnarray}
S_R^{(2)}=\xi\ln \tilde{t}
\end{eqnarray}
All is shown explicitly in Fig.~\ref{Fig:OTOC}(c) and (e). Finite size scaling is testified in Fig.~\ref{Fig:OTOC}(e), and a volume law for saturate R\'{e}nyi entropy could be easily seen.

From all these we can see some connections between logarithmic light-cone, butterfly velocity and entanglement spreading. A rule can be guessed as mutual information is formed and formed only in ``information light-cone".

\emph{Quantum recurrence estimation} In above calculations we assumed that quantum recurrence does not happen, now we estimate quantum recurrence time. To estimate the lower bound of quantum recurrence time, we will first alter the interaction law for distance. Later we will extend our calculation to exponential decay interaction. Now we assume $J_{ij}=J_{ij}^0/|i-j|$. Instead of taking $J_{ij}^0$ random distributed, we assume $J_{ij}^{0}=\pm J(J>0)$ for equal probability. Then from Eqn.(\ref{Eq:ND}) we know to make $f(t)$ 1 again, we have to make all period in $f(t)$ meet each other. Therefore the large period should be the multiple of all small periods. 
For instance, all possible periods are $(2n\pi/4J )(n=1,2,3\cdots, N)$. Let us denote common multiple of $1,2,\cdots,N$ as ${\cal M}_N$,
\begin{eqnarray}
{\cal M}_N= 2^{\frac{\ln N}{\ln 2}}3^{\frac{\ln N}{\ln 2}}5^{\frac{\ln N}{\ln 5}}\cdots p^{\frac{\ln N}{\ln p}}=e^{\pi(N)\ln N},
\end{eqnarray}
where the product is for all prime numbers less than N, p is the largest prime number, $\pi(N)$ is number of prime numbers less than N.  Approximately $\pi(N)\approx Li(N)=N/\ln N+N/(\ln N)^2+\cdots$. Therefore $\ln {\cal M}_N=Li(N)\ln N\approx N$, that is, the quantum recurrence time for such system is of $e^N$ order. Then in a similar way for $J_{ij}=J_{ij}^0/|i-j|^\alpha$, the quantum recurrence time must be larger than $e^{N^\alpha}$. If $J_{ij}$ is exponentially small, we have to calculate the common multiple less than $e^N$, that makes $e^{e^N}$, therefore recurrence could hardly happen. Further, random value for $J_{ij}^0$ makes the recurrence time even longer, therefore it justified our assumption.

From the calculation of quantum recurrence, we could also see the decay of OTOC $f(t)$ come from dephasing of incommensurate frequencies, while quantum recurrence happens because of resonances of different periodicity. 
  
\emph{Discussion} In this Letter, we carried out an analytical calculation of OTOC for a typical MBL system. We find an universal power law decay behavior of OTOC between the LLC arrival time $\tau_B$ and the saturation time $\tau_S$ as $2^{-\xi\ln t}$, 
showing logarithmic slow scrambling distinct from zero scrambling behavior in AL and exponential fast scrambling in quantum chaos. The dimensionless localization length $\xi$ is found to be both the R\'{e}nyi entropy growth rate and the Lieb-Robinson velocity. Experimentally we could measure the R\'{e}nyi entropy from the input and output of our quantum circuit by a two chain setup\cite{Qi}. We also noticed a recent spin echo spectrum measurement proposal\cite{DEER}, where the DEER response could capture very similar information without reversing hamiltonian in evolution. But still the DEER response is not OTOC, therefore the similarity and difference between DEER response and OTOC are still puzzling us. Inspired by DEER response, noticing that the $\sigma^{0}$, $\sigma^3$ are similar and $\sigma^1$, $\sigma^2$ are similar, we could choose $\hat{W}=R_A^{\pi/2}=\prod_{i\in A}\frac{1}{\sqrt{2}}(1-i\sigma_i^y)$ and $\hat{V}=R_D^{\pi/2}=\prod_{i\in A}\frac{1}{\sqrt{2}}(1-i\sigma_i^y)$, then the summation over different routes will be automatically done. Then this can be measured by interference experiment by recent proposals for OTOC measurement in cold atom systems with the help of cavity or ancilla \cite{OTOC_proposal00,OTOC_proposal01,OTOC_proposal02}. Other plans as measuring entanglement entropy by quantum revival serves also as an indirect test for OTOC\cite{Moore15}.

There are still a lot of future studies on going like finite temperature effect and the OTOC behavior across MBL-ETH transition, therefore we could expect the OTOC will bring us more insights about information scrambling in general systems in the future.

\emph{Note} Recently, there are four other works in the same direction appearing on arXiv. To be time-ordered, they are works by Xie Chen's group\cite{Xie}, Hui Zhai's group\cite{Hui}, Brian Swingle and Debanjan Chowdhury\cite{Brian16}, Rongqiang He and Zhongyi Lu\cite{He}.

\emph{Acknowledgement} Y. C would like to thank Yingfei Gu, ChaoMing Jian, Ruihua Fan and Chushun Tian for inspiring discussions.

\end{document}